\documentstyle[prl,aps]{revtex}
\input{psfig.sty}
\draft
\begin{document}
\twocolumn[\hsize\textwidth\columnwidth\hsize\csname@twocolumnfalse\endcsname
\title{Dynamics of  axial separation  in  long rotating drums}
\author{I. Aranson}
\address{
Argonne National Laboratory,
9700 South  Cass Avenue, Argonne, IL 60439}
\author{L.S. Tsimring}
\address{Institute for Nonlinear Science, University of California, San Diego, La Jolla,
CA 92093-0402
}
\date{\today}
\maketitle
\begin{abstract}
We propose a continuum  description for the axial separation of granular
materials in a long rotating drum. The model,  operating  with two
local variables, concentration difference and the dynamic angle of
repose, describes both initial transient traveling wave dynamics and
long-term segregation of the binary mixture. Segregation proceeds
through ultra-slow logarithmic coarsening. 
\end{abstract}

\narrowtext
\vskip1pc]

The collective  dynamics of granular materials recently have been
attracting much interest \cite{nigel,swinney,hill,drum,zik,morris}.
Intrinsic dissipative nature of interaction among macroscopic particles
sets granular matter  apart from familiar gaseous, liquid, or solid
states. One of the most fascinating features of heterogeneous granular
materials is their ability to segregate under external agitation instead
of mixing as would be expected from thermodynamics.  Essentially any
variation in mechanical properties of particles (size, shape, density,
surface roughness) may lead to their segregation.   Long rotating drums
partially filled with a mixture of grain sizes exhibit both radial and
axial size segregation \cite{hill,morris}. In radial segregation grains
of one type (for grains of different sizes, the smaller ones) build up a
core near the axis of rotation.  Axial segregation often follows radial
segregation, with the mixture of grains separating into bands according
to size arranged along the axis of the drum.  Axial segregation leads to
either a stable array of concentration bands, or to the complete
segregation \cite{chicarro}.  

Granular flow in a rotating drum is  different from conventional fluid
motion.   In the bulk, particles perform solid body rotation around the
drum axis until they reach the free surface. Then they slide down within
a thin near-surface layer\cite{drum}. The surface has a $S$-curved shape
and its average slope is determined by the {\em dynamic angle of
repose}.   Since there is almost no inter-particle motion in the bulk,
the segregation predominantly  occurs within the fluidized near-surface
layer.  The radial segregation occurs during first few revolution of the
drum.   For long drums (length much larger than radius), along with
radial segregation, {\em axial segregation} occurs at much later stages
(after hundreds of revolutions). Recent
experiments\cite{hill,zik,morris} revealed interesting features of axial
segregation. At early stages, small-scale perturbations travel across
the drum in both directions (it was obvious from the dynamics of
pre-segregated mixtures\cite{morris}), while at later times more
long-scale static perturbations take over and and lead to
quasi-stationary bands of separated mixture. Bands of segregated
materials interact at a very long time scale and exhibit logarithmic
coarsening\cite{zik}.  This process can be accelerated in a drum of a
helicoidal shape.  The bands can be locked  in a drum with the radius
modulated along the axis \cite{zik}.

Most of the theoretical models of segregation agree in that the
underlying reason for segregation is the sensitive dependence of the
surface slope or shape on the relative concentration of different
particles in the mixture\cite{zik}.   In Ref. \cite{zik} a simple theory
was proposed of  segregation in thin surface flow driven by the local
slope, combined with conservation of particles. For monodisperse
material, the model recovers the $S$-shape of the free surface.   For a
binary mixture the model yields a nonlinear diffusion equation for the
relative concentration of  the ingredients along  horizontal axis.
Axial segregation occurs when the diffusion coefficient turns negative.
This model yields a significant insight into the nature of the
instability leading to the segregation, however since it is based on a
first-order diffusion equation, it fails to describe the traveling waves
observed at the early stages of axial segregation\cite{morris}. 

In this Letter  we propose a continuum model which describes
consistently the early phase of segregation with traveling waves as well
as the later stage of segregation characterized by  slow merging of
bands of different particles. Our model predicts slow (logarithmic)
coarsening of the segregated state. The dynamics of segregation shows
striking similarity with the experiments of Ref. \cite{morris}. 

Let us consider a  mixture of two sorts of particles, $A$ and $B$, of
which $A$($B$) corresponds to particles with larger (smaller) static
repose angle.   Our model operates with two continuum  variables:
the  relative concentration of particles $c=(c_A-c_B)/\langle c
\rangle$, and local dynamic repose angle $\theta$.  Here $c_{A,B}$ are
local partial concentrations of particles, and $\langle c \rangle
=\langle c_A+c_B \rangle $ is an average over whole system total
concentration.   We assume that $c$ and $\theta$ are functions of
longitudinal coordinate $x$ and time $t$. In fact, concentration also
depends on the radial coordinate, however in this study we are concerned
with axial segregation, and will operate with quantities integrated over
cross-sections of the long drum.

The first equation represents conservation of the relative concentration 
$c$ in the binary mixture: 
\begin{eqnarray} 
\partial_t c=-\partial_{x}(-D\partial_{x} c + g(c)\partial_{x}\theta) \label{conc}
\end{eqnarray} 
The first term describes diffusion flux (mixing), and the second term
describes differential flux of particles due to the gradient of dynamic
repose angle.  Similar term with specific function $g(c)=G_0(1-c^2)$ (in
our notation) was first derived in Ref. \cite{zik} from the condition of
detailed  flux balance.  The proportionality constant $G_0$ depends on
the physical properties of the grains (see \cite{zik}).  For simplicity,
the constant $G_0$ can be eliminated by the scaling of distance $x \to
x/ \sqrt{G_0}$.   The sign $+$  before this term means that the
particles with the larger  static repose angle are driven towards
greater dynamic repose angle. As will be shown below, this differential
flux gives rise to the segregation instability. Since this segregation
flux vanishes with $g(c)$ $|c|\to 1$ (which correspond to pure $A$ or
$B$ states), it provides a natural saturation mechanism for the
segregation instability. 

The second equation describes the dynamics of $\theta$ 
\begin{eqnarray}
\partial_t\theta&=\alpha(&\Omega - \theta + f(c)) + 
D_\theta\partial_{xx}\theta + \gamma\partial_{xx} c.\label{theta}
\end{eqnarray}
Here $\Omega$ is the angular velocity of the drum rotation, and $f(c)$
is the static angle of repose which depends on the relative
concentration.  According to our definition of $c$, $f(c)$ is an
increasing function of relative concentration\cite{hill1}. Since the
angle of repose as a function of the concentration $c$ changes typically
in a small  range, we can approximate the function $f(c)$ by linear
dependence $f(c)= F + f_0 c  $. The constant $F$ can be eliminated by
the substitution  $\theta \to \theta -F$.   First term in the r.h.s. of
Eq.(\ref{theta}) describes the local dynamics of the repose angle
($\Omega$ increases the angle, and $-\theta+f(c)$ describes the
equilibrating effect of the surface flow), and the term $
D_\theta\partial_{xx}\theta$ describes axial diffusive relaxation.   The
last term, $ \gamma \partial_{xx}c$, represents the lowest-order
non-local contribution from the inhomogeneous distribution of $c$ (the
first derivative $\partial_x c$ cannot be  present  due to reflection
symmetry $x\to -x$). As we will see later, this term gives rise to the
transient oscillatory dynamics of the binary mixture. 

The stationary uniform state of the system is $c=c_0; \theta_0=\Omega+f_0
c_0$ where $c_0$ is determined by initial conditions.  $\theta_0$ has a
meaning of the stationary dynamic repose angle, which in the limit
$\Omega\to 0$ approaches the static repose angle\cite{note}.  

\begin{figure}
\centerline{\psfig{figure=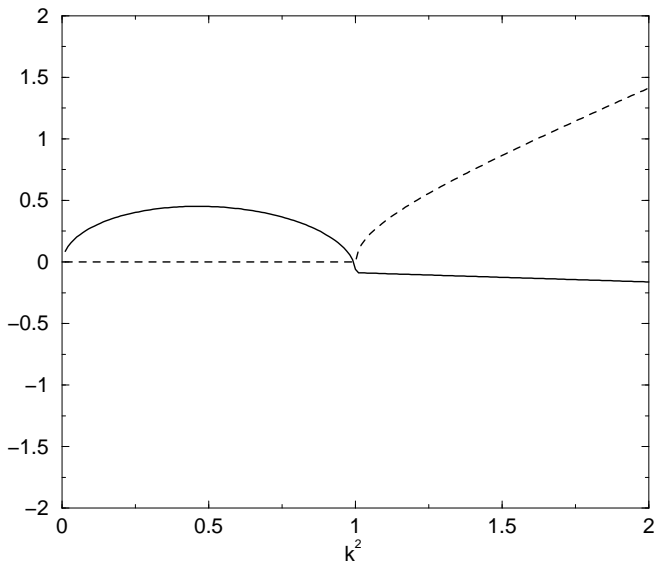,height=3.in}}
\caption{Dispersion relation $\lambda(k)$ for the small perturbation of
the uniform state $c_0=0$ at $f_0=40$, $D=0.05$,$ D_\theta=0.1$,
$\alpha=0.025$, $\gamma=1$. Perturbations are
unstable at $k<k_c=0.994$ and oscillate (and decay) at $k>k_*=1$.}
\label{fig1}
\end{figure}

Let us consider the stability of the  uniform state,
\begin{eqnarray}
c=c_0+Ce^{\lambda t +ikx},\nonumber\\
\theta=\theta_0+\Phi e^{\lambda t +ikx}
\end{eqnarray}
To linearize the system, we need to expand the function $g(c)$ near
$c_0$. The stability properties depend on the values of $f_0$ and  
$g_0\equiv g(c_0)$. The eigenvalues $\lambda_{1,2}$ 
are found from the following equation,
\begin{equation}
(\lambda+D k^2)(\lambda+D_\theta k^2+\alpha)-g_0(\alpha*f0 k^2-\gamma k^4)=0.
\label{eigen}
\end{equation}
At  $k \to 0 $  one finds $\lambda_1=-\alpha-(D_\theta+g_0f_0 )k^2; 
\ \lambda_2=(g_0f_0 -D)k^2$.
Asymptotic expansion at $k \to \infty$  yields
\begin{equation}
\lambda_{1,2}=\frac{1}{2}\left(-D_\theta-D\pm\sqrt{(D_\theta-D)^2-4g_0\gamma}\right)k^2.
\end{equation}
It is easy to see that if $g_0f_0 >\alpha D$, long-wave perturbations
are unstable ($\lambda_2>0$), and if $g_0\gamma>(D_\theta-D)^2/4$,
short-wave perturbations oscillate and decay (eigenvalues
$\lambda_{1,2}$ are complex conjugate with negative real part). The full
dispersion curve $\lambda(k^2)$ for particular values of parameters is
plotted in Fig. \ref{fig1}.  This curve is consistent with the
measurements in Ref. \cite{morris}.  Indeed, at small wavelengths the
perturbations travel, and their phase velocity $v_{ph}={\rm Im}
\lambda/k$ grows linearly with $k$. At small $k$, frequency and phase
velocity turn into zero and remain zero for all long-wave perturbations
$k<k_*$. Perturbations in the range $0<k<k_c$ where $k_c\equiv(\alpha(g_0f_0
-D)/(g_0\gamma+DD_\theta))^{1/2}<k_*$, grow exponentially and there
is an optimal wavenumber of the fastest growing perturbations $k_0$.
Oscillating perturbations with $k>k_*$ always decay, which seems to be
in agreement with experimental data, although direct experimental
measurement of ${\rm Re} \lambda(k)$ is lacking.

\begin{figure}
\centerline{a) \psfig{figure=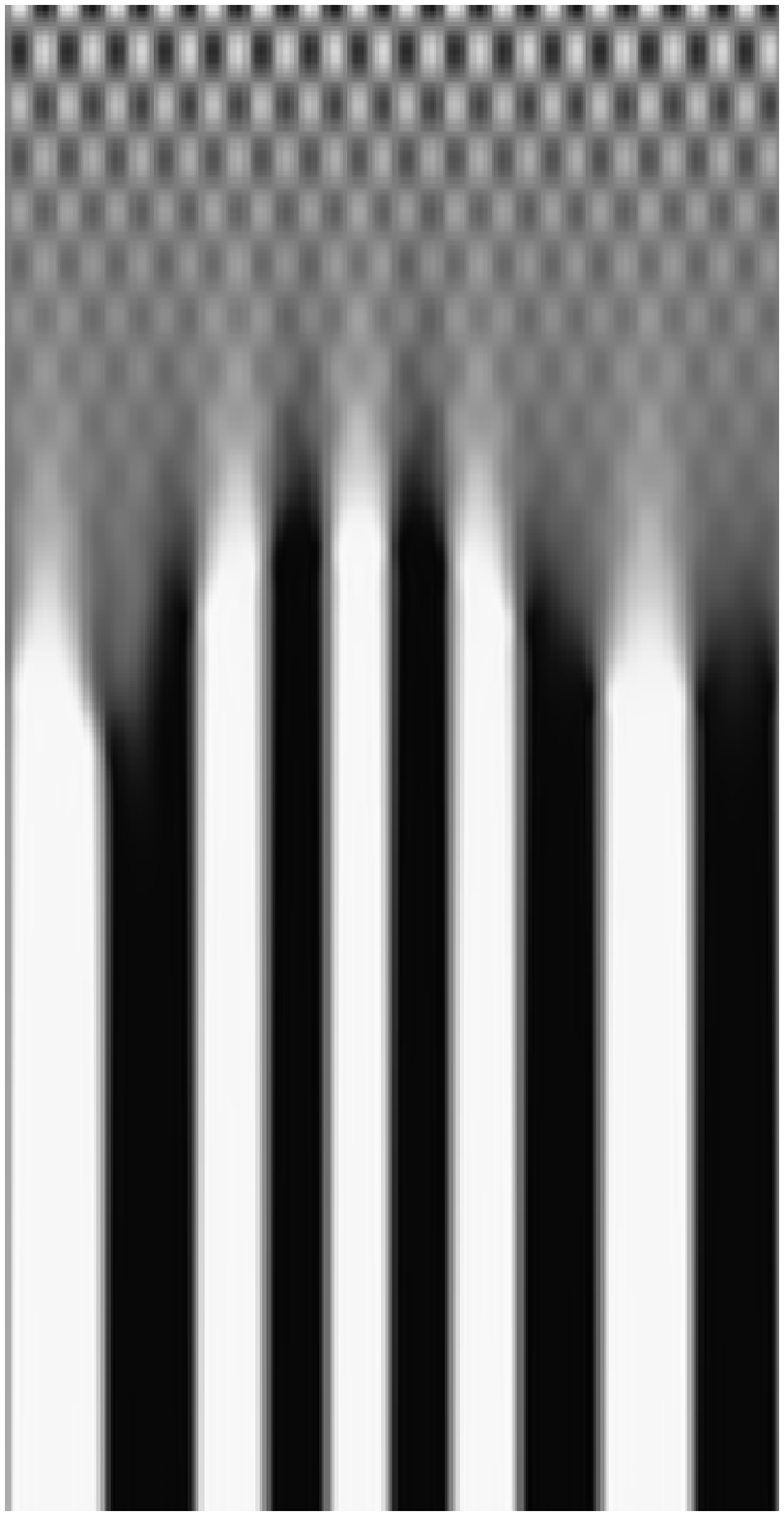,height=3in} 
\hspace{1.0cm} b) \psfig{figure=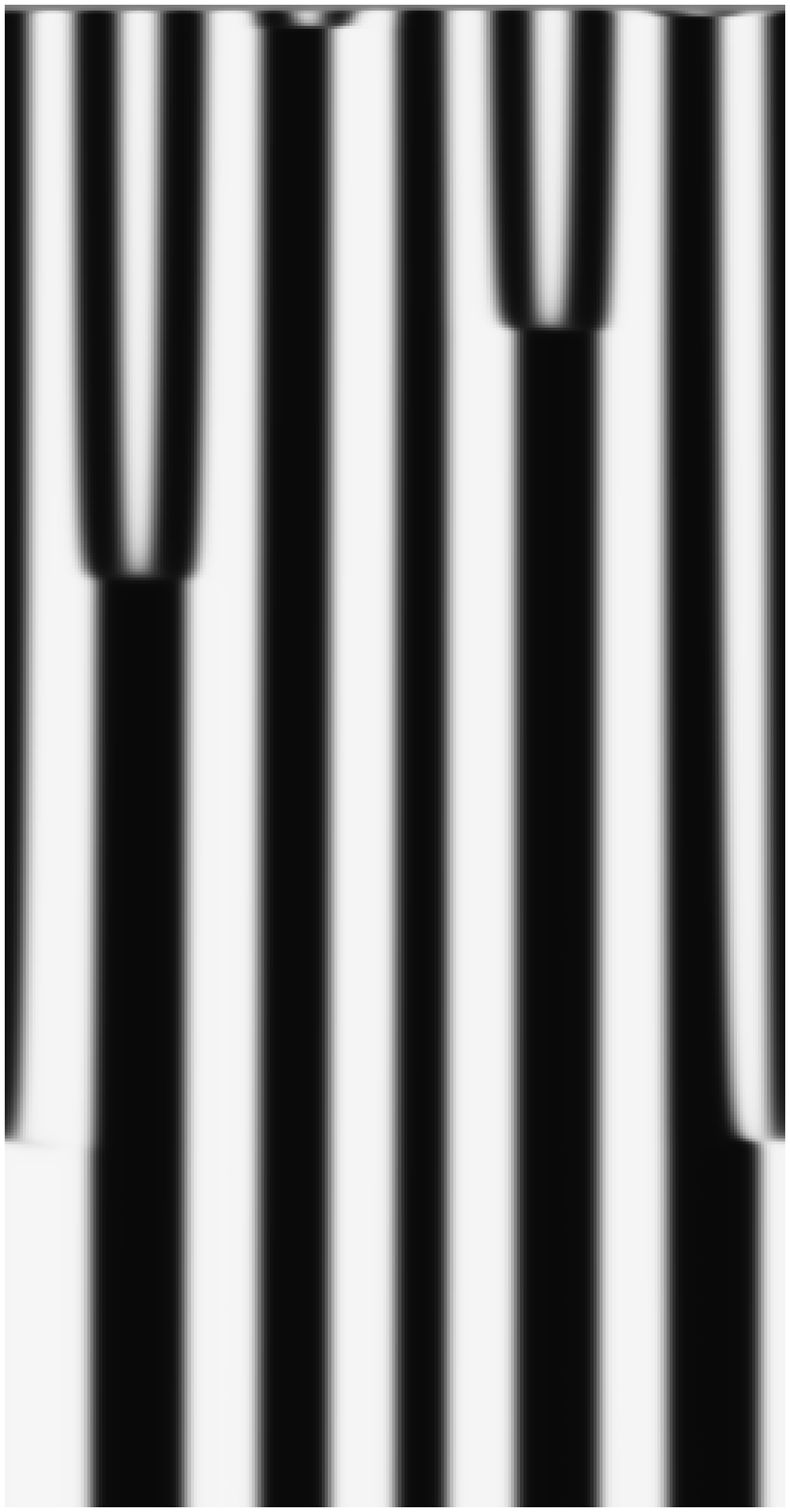,height=3in}  }
\caption{Space-time diagrams of the evolution of the pre-separated 
state ($c(x,0)=c_0+c_i\cos(k_0 x)$) with $c_0=0.,\ c_i=0.95,$ and
$k_0=1.79>k_*$);{\em a} initial transient, 
At times $t<15$ the initial perturbations excite decaying
standing wave (superposition of left- and right-traveling waves), and 
at larger times $t>15$, aperiodic segregated bands emerge. 
Parameters of the model are: 
$D=0.05, \gamma=1, \alpha=0.025, f_0=40, L=60$ and $ D_\theta=0.1$. {\em b},
Space-time diagram for long-time evolution, band merging and coarsening
during long-time evolution ($0<t<10000$) at higher diffusion constants, parameters
$D=0.8,\ D_\theta=0.5,\ \gamma=1,\ \alpha=0.5,\ f_0=2,\ L=140$.}

\label{fig2}
\end{figure}

\begin{figure}
\centerline{\psfig{figure=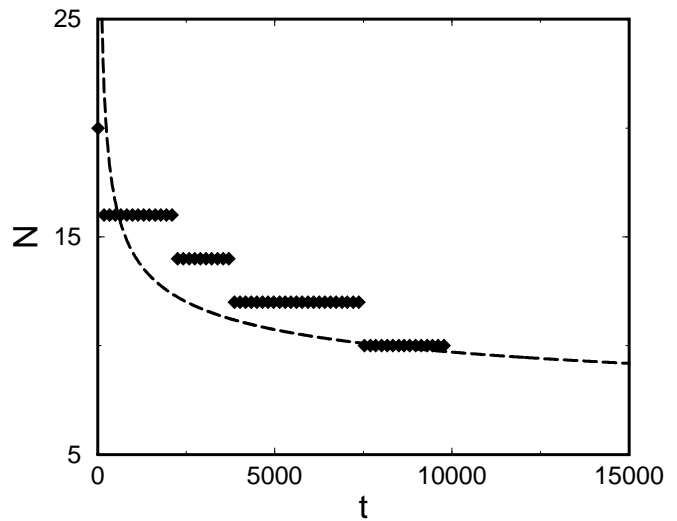,height=3.in}}
\caption{Number of fronts $N$ as a function of time (diamonds) and its fit
by a function $N=70/(\ln t - 2.5 )$ (long-dashed line). Parameters correspond 
to Fig.2b}
\label{fig3}
\end{figure}

We performed numerical simulations of Eqs.(\ref{conc}-\ref{theta}) using
pseudo-spectral split-step method with periodic  boundary conditions.
We used up to 512 mesh points in our numerical procedure.  The following
form of nonlinear functions in Eqs. (\ref{conc},\ref{theta}) was
implemented: $g(c)=1-c^2$ and $f(c)=f_0 c$ \cite{note1}.
The dynamics of the initially pre-separated state with wavenumber
$k>k_*$ in a system with size $L=60$ is shown in Fig. \ref{fig2}. In
agreement with experiments\cite{morris}, short-wave initial perturbation
produce decaying standing waves, which later are replaced by
quasi-stationary  bands (Fig. \ref{fig2}a). The bands are separated by
sharp interfaces which are very weakly attracted to each other. In fact,
in simulations with parameters corresponding to Fig.\ref{fig2}a we were
not able to detect interface merging at all in a reasonable 
simulation time.  However at higher rates of diffusion and dissipation,
corresponding to different experimental conditions (rotation frequency,
filling factor, etc), the interaction becomes more significant, and it
leads to band merging and overall pattern coarsening (see Fig.
\ref{fig2}b). In Fig. \ref{fig3}, we present a number of bands as a
function of time for this run.

Fronts separating bands of different grains, can be found as
stationary solutions of Eqs.(\ref{conc}),(\ref{theta}). 
In an infinite system one finds from stationary Eq.(\ref{conc})
$\theta=\theta_0+DG(c)$, where $G(c)=\int[g(c)]^{-1}dc=-\frac{1}{2} \log
\frac{1-c}{1+c}$ and $\theta_0$ is an integration constant. Plugging this 
expression in Eq.(\ref{theta}), we obtain the second order differential 
equation for $c$ (for a symmetric solution one chooses $\theta_0=\Omega$),
\begin{eqnarray}
\frac{d}{dx}\left[\left(\gamma+
\frac{D D_\theta}{1-c^2}\right)\frac{dc}{dx}\right]+  
 \alpha f_0 c+ \frac{ \alpha D}{2} \log \frac{1-c}{1+c}=0.
\end{eqnarray}
It is easy to see that this equation possesses an interface solution.
The asymptotic behavior of this solution can be found in the limit $ D
\ll f_0 $, when the states on both sides of the interface are well
segregated ($|c(x\to\infty)|\to \pm 1)$. In this limit, far away from
the interface $1-|c|\propto \exp(-x/d_0)$ where $d_0=
\sqrt{\frac{D_\theta}{\alpha} +\frac{ 4 \gamma}{\alpha   D} \exp (
-\frac{2f_0}{ D} )}  $. As seen from this formula, the characteristic
front width vanishes as $D,D_\theta\to 0$.  This result could be
anticipated, as in the absence of diffusion nonlinearity $g(c)$ drives
the system towards complete segregation.

Slow logarithmic coarsening of the segregated state can be understood in
terms of the weak interaction of fronts.   Since the asymptotic field of
the front approaches equilibrium value of the concentration
exponentially fast, we expect in general exponentially weak interaction
between the neighboring fronts leading to logarithmic times for the
front annihilation.  

The problem of front interaction has been
recently considered\cite{fraerman} for the Cahn-Hilliard model which
arises in relaxation phase ordering kinetics\cite{CH}. As in the Cahn-
Hilliard model, the order parameter (here concentration $c$) is a
conserved quantity, therefore front interaction must conform to a global
constraint. In fact, in a particular case $g(c)=1,\ D=0,\
D_\theta=0$ our model (\ref{conc}),(\ref{theta}) can be reduced 
to a single equation for the relative concentration
\begin{equation}
\partial_{tt}c+\partial_t c=-\partial_{xx}(f(c)+\partial_{xx}c)
\label{ch}
\end{equation}
which differs from the Cahn-Hilliard model only by the first term in the
l.h.s. This term describes non-potential effects, including traveling
waves discussed above. For the slow process of the band relaxation in
the long-time limit this term can be dropped.  Front solutions within
the Cahn-Hilliard equation exist for $N$-shaped functions $f(c)$, for
example, a cubic polynomial $f(c)=2(c-c^3)$. In this case stationary
interface solution is found analytically as $c(x)=c_0+\tanh(x-x_0)$, and
interaction of the interfaces can be analyzed. This
analysis\cite{fraerman}, predicts that a single band of positive $c$ in
the sea of negative $c$ (two fronts) can either annihilate or reach a
stationary width depending on the initial distance between fronts. If a
number of fronts is greater than 2, front interaction leads to their
annihilation and pattern coarsening. The interaction between fronts due
to their tail overlapping is exponentially weak. The rate of front
interaction can be determined using multiple scale analysis. In the
simplest case of three equidistant fronts, two outer fronts collide with
the central front and disappear after time $T\propto d\exp(d/d_0)$ where
$d$ is the initial distance between fronts. Thus, for multi-band
structure, the number of fronts  $N$ (proportional to the inverse
average distance between fronts) decreases approximately  as a
logarithmic function of time ($N =1/d \sim 1/(const + d_0\log T) $.
Since interfaces in our more complicated model have similar
exponentially small tails, we anticipate a similar logarithmic law of
band coarsening. This dependence indeed agrees with our numerical
simulations (see Fig. \ref{fig3}).

In conclusion, we proposed a simple continuum model for axial
segregation of binary granular mixtures in long rotating drums. The
model operates with two local dynamical variables, relative
concentration of two components and dynamic repose angle.   The dynamics
of our model shows qualitative similarity with the experimental
observations of initial transients and long-term segregation
dynamics\cite{zik,hill,morris}.   It captures both initial transient
traveling waves and subsequent onset band structure. The dispersion
relation for slightly perturbed uniform state (\ref{eigen})
qualitatively agrees well with observations\cite{morris}, and can serve
for fitting the model parameters.  The logarithmic coarsening of the
quasi-static band structure which follows from our model, deserves
experimental verification. This coarsening is typical for systems with
exponentially weak attractive interaction among defects or interfaces,
as in the phase ordering kinetics described by the Cahn-Hilliard model.
Our simulations also showed that the model qualitatively reproduces more
complicated phenomenology of the separation process reported in
Ref.\cite{zik}.   In particular, periodic modulation of the drum radius,
modeled in our approach by periodic variation of $\Omega$, leads to band
locking.  Breaking of the $x \to -x$ symmetry by term $\partial_x \theta$,
introduced in r.h.s. of Eq.   (\ref{conc}), results in complete
segregation, similarly  to the dynamics of grains in the drum with
helicoidal shape. 

We thank S. Morris and J. Kakalios for useful discussions. This research
is supported by the US Department of Energy, grants \#
W-31-109-ENG-38, DE-FG03-95ER14516, DE-FG03-96ER14592, and by NSF, 
STCS \#DMR91-20000.


\end{document}